\documentclass[letterpaper,10pt]{article}

\usepackage{amsmath}
\usepackage{amssymb}
\usepackage{psfrag}
\usepackage[dvips]{graphicx}
\usepackage{epsfig}

\newcommand{\dirac}{{\slash \negthinspace \negthinspace \negthinspace \nabla}}
\newcommand{\dd}{\textrm{d}}

\author{
A. L\'opez-Ortega\thanks{alopezo@ipn.mx} \\
Departamento de F\'{\i}sica. Escuela Superior de F\'{\i}sica y Matem\'aticas. \\
Instituto Polit\'ecnico Nacional. \\
Unidad Profesional Adolfo L\'opez Mateos, Edificio 9. \\
M\'exico, D.\ F., M\'exico. \\
C.\ P.\ 07738 \\ 
and\\
Centro de Investigaci\'on en Ciencia Aplicada y Tecnolog\'{\i}a Avanzada. \\ 
Unidad Legaria. Instituto Polit\'ecnico Nacional. \\
Calzada Legaria \# 694. Colonia Irrigaci\'on. Delegaci\'on Miguel Hidalgo. \\
M\'exico, D.\ F., M\'exico. \\
C.\ P.\  11500  
}

\title{Classical stability of black holes under massless Dirac perturbations}

\begin{document}

\maketitle

\begin{abstract}
 
In a $D$-dimensional maximally symmetric spacetime we simplify the massless Dirac equation to two decoupled wavelike equations with effective potentials. Furthermore in $D$-dimensional Schwarzschild and  Schwarzschild de Sitter black holes we note that for the massless Dirac field moving in the region exterior to the event horizon at least one of the effective potentials is not positive definite. Therefore the classical stability of these black holes against this field is not guaranteed. Here with the help of the $S$-deformation method, we state their classical stability against the massless Dirac field, extend these results to maximally symmetric black holes, and comment on the applicability of our results to establish the stability with respect to other classical fields. \\

KEYWORDS: Classical stability; Dirac field; Higher dimensional black holes

\end{abstract}

\section{Introduction}
\label{sec: introduction}

The analysis of the dynamics of the classical fields in black hole spacetimes is necessary to calculate significant physical quantities and in the study of relevant physical phenomena \cite{Kokkotas:1999bd}--\cite{Konoplya:2011qq}, for example, by investigating the dynamics of perturbing fields we can determine the classical stability of the black holes with respect to linear perturbations  \cite{Regge:1957td}--\cite{Cardoso:2010rz}.

Recently the higher dimensional black holes have attracted attention (for example see Ref.\  \cite{Emparan:2008eg}), and for several four and higher dimensional spherically symmetric black holes their classical stability against gravitational and electromagnetic perturbations is studied \cite{Regge:1957td}--\cite{Cardoso:2010rz}. We think that a relevant contribution to the analysis of gravitational theories is to determine the classical stability of the black holes with respect to the small perturbations.

To show the stability of a static spherically symmetric black hole against a linear perturbation a frequently used method is to prove that its dynamics is governed by a positive self-adjoint operator or by one that can be extended to a positive self-adjoint operator \cite{Ishibashi:2003ap}--\cite{Ishibashi:2011ws}. In the following sections we shall use this method. 

As is well known, in curved spacetimes the Dirac field behaves in a different way than the boson fields, for example, this fermion field does not suffer superradiant scattering in rotating black holes \cite{Martellini:1977qf,Unruh:1973,Iyer:1978du}. Furthermore the behavior of Dirac fields is not explored as extensively as for boson fields \cite{Kokkotas:1999bd}--\cite{Konoplya:2011qq}. Thus it is convenient to put attention to the dynamics of fermion fields in higher dimensional black holes. 

We are aware of Refs.\ \cite{Gibbons:1993hg}--\cite{Wu:2008bc} in which the behavior of the Dirac field is analyzed in higher dimensional backgrounds. As far as we know at present time is not solved the question of the classical stability of higher dimensional black holes against Dirac perturbations. To partially answer this issue, here for the massless Dirac field propagating in maximally symmetric black holes we simplify its equations of motion to a decoupled pair of wavelike equations with effective potentials. In the $D$-dimensional Schwarzschild and Schwarzschild de Sitter (SdS) spacetimes we notice that the effective potentials are not positive definite, as a consequence we can not guarantee the positivity of the operator that governs the dynamics, and hence it is not straightforward to show the classical stability of these black holes with respect to the massless Dirac perturbations. We find similar examples in other maximally symmetric black holes.

Although we expect that the $D$-dimensional maximally symmetric black holes are stable under massless Dirac perturbations, we believe that it is necessary to prove that these fermion fields do not produce instabilities. Hence in what follows we state the classical stability of several higher dimensional black holes against massless Dirac perturbations.

We organize the rest of the paper as follows. For $D$-dimensional maximally symmetric spacetimes in Sect.\ \ref{sec: Massless Dirac equation}  we simplify the massless Dirac equation to a pair of decoupled wavelike equations with effective potentials. In Sect.\ \ref{sec: Effective potentials in} we analyze and plot the effective potentials of the massless Dirac field propagating in Schwarzschild and  SdS black holes. We find that in these two spacetimes at least one effective potential is not positive definite. In Sect.\ \ref{sec: Stability analysis} we establish the classical stability of the $D$-dimensional Schwarzschild and SdS black holes under massless Dirac perturbations and we extend these results to the $D$-dimensional maximally symmetric black holes. Finally in Sect.\ \ref{sec: Discussion} we discuss our main results and comment on the application of these to prove the classical stability with respect to other perturbing fields.

\section{Massless Dirac equation}
\label{sec: Massless Dirac equation}

As is well known, in a $D$-dimensional ($D \geq 4$) spacetime with line element of the form
\begin{equation} \label{eq: maximally symmetric metric}
\dd s^2 = F(r)^2 \dd t^2 -G(r)^2 \dd r^2 - H(r)^2 \dd \Sigma^2_{D-2},
\end{equation} 
where $\dd \Sigma^2_{D-2}$ is the line element of a $(D-2)$-dimensional invariant base manifold, the Dirac equation
\begin{equation} \label{eq: Dirac equation}
 i \dirac \psi = m \psi ,
\end{equation} 
simplifies to the coupled system of partial differential equations \cite{Gibbons:1993hg,Das:1996we,LopezOrtega:2004cq,LopezOrtega:2009qc}
\begin{align} \label{eq: Dirac equation simplified}
 \partial_t  \psi_2 - \frac{F}{G} \partial_r  \psi_2 & =  \left( i \kappa \frac{F}{H} -  i m F \right) \psi_1,  \\
 \partial_t \psi_1 + \frac{F}{G} \partial_r \psi_1 & =  - \left( i \kappa \frac{F}{H} +  i m F \right) \psi_2,  \nonumber
\end{align} 
where  $\psi_1$, $\psi_2$ are the components of a two-dimensional spinor in the $(t,r)$ sector of the metric (\ref{eq: maximally symmetric metric}), and $\kappa$ denotes the eigenvalues of the Dirac operator on the $(D-2)$-dimensional base manifold $\dd \Sigma^2_{D-2}$, that is, $\dirac_{\dd \Sigma} \chi = \kappa \chi$.\footnote{Notice that in Eqs.\ (30) of Ref.\ \cite{LopezOrtega:2009qc}, the factor $-i m F \psi_1$ ($-i m F \psi_2$) in the right hand side of Eqs.\ (\ref{eq: Dirac equation simplified}) is erroneously written as $-i \mu F \psi_1$ ($-i \mu F \psi_2$), that is, in those equations of Ref.\ \cite{LopezOrtega:2009qc} the mass $m$ is denoted by $\mu$ without justification.} In what follows we call to a spacetime with line element of the form (\ref{eq: maximally symmetric metric}) as maximally symmetric black hole.

For the massless Dirac field, if we define
\begin{equation} \label{eq: W function general}
 W = -i \kappa \frac{F}{H},
\end{equation}
then we rewrite Eqs.\ (\ref{eq: Dirac equation simplified}) as
\begin{align} \label{eq: coupled equations Zplus Zminus}
\partial_x Z_+ - \partial_t Z_- &= W Z_+ , \\
\partial_x Z_- - \partial_t Z_+ &= - W Z_- , \nonumber
\end{align} 
with 
\begin{equation} \label{eq: definitions Z and tortoise}
 Z_\pm = \psi_2 \pm \psi_1, \qquad \qquad \textrm{and} \qquad \qquad \frac{\dd x}{\dd r} = \frac{G}{F},
\end{equation} 
that is, $x$ is the tortoise coordinate of the spacetime (\ref{eq: maximally symmetric metric}). 

From Eqs.\ (\ref{eq: coupled equations Zplus Zminus}) we obtain the wavelike equations for the functions $Z_\pm$
\begin{equation}
 \partial_x^2 Z_\pm - \partial_t^2 Z_\pm = ( W^2 \pm \partial_x W ) Z_\pm ,
\end{equation} 
that is
\begin{equation} \label{eq: wavelike equations Dirac}
 \partial_t^2 Z_\pm = \partial_x^2 Z_\pm - V_\pm Z_\pm ,
\end{equation} 
where 
\begin{equation} \label{eq: potentials plus minus}
 V_\pm = W^2 \pm \partial_x W .
\end{equation} 

In what follows we call to the quantities $V_\pm$ the effective potentials and these are usually obtained for Schr\"odinger type equations after we take a harmonic time dependence \cite{Chandrasekhar book}, here we show that for the massless Dirac field a similar procedure works for the coupled partial differential equations (\ref{eq: Dirac equation simplified}). For the massive Dirac field we do not get a similar simplification. Although for the massless Dirac equation the previous reduction is straightforward, it may be useful to study with numerical methods the behavior of the massless Dirac field in curved spacetimes (see for example Ref.\ \cite{Gundlach:1993tp}).

\section{Effective potentials in Schwarzschild and SdS black holes}
\label{sec: Effective potentials in}

The metrics of the $D$-dimensional Schwarzschild and SdS black holes take the form (\ref{eq: maximally symmetric metric}), where $\dd \Sigma^2_{D-2}$ is the line element of the $(D-2)$-dimensional unit sphere. Therefore the eigenvalues $\kappa$ take the form \cite{Camporesi:1995fb}
\begin{equation}
\kappa = \pm i \left(l + \frac{D-2}{2} \right),                                                                                                                                                                                                                                                                                                                                                                                                                                                                                                                                                                                                      \end{equation} 
where $l=0,1,2,\dots$ In what follows we consider only the eigenvalues $\kappa = i \left(l + (D-2)/2 \right)$. We think that for the eigenvalues $\kappa = - i \left(l + (D-2)/2 \right)$ we find equivalent results.\footnote{To obtain the effective potentials for the eigenvalues $\kappa = - i \left(l + (D-2)/2 \right)$, we must change the labels $+ \leftrightarrow -$ in the effective potentials $V_+$ and $V_-$ of the eigenvalues $\kappa = i \left(l + (D-2)/2 \right)$.}

For the $D$-dimensional Schwarzschild and SdS black spacetimes we know that the metric functions of the line element (\ref{eq: maximally symmetric metric}) satisfy
\begin{equation} \label{eq: identifications metric functions}
 F^2 = \frac{1}{G^2} = f, \qquad \qquad H= r,
\end{equation} 
where 
\begin{equation} \label{eq: metric function Schwarzschild}
 f = 1 - \frac{2 \mu}{r^{D-3}},
\end{equation} 
for the $D$-dimensional Schwarzschild black hole and
\begin{equation} \label{eq: metric function Schwarzschild de Sitter}
 f = 1 - \frac{2 \mu}{r^{D-3}} - \lambda r^2,
\end{equation} 
with $\lambda > 0$ for the $D$-dimensional SdS black hole ($\lambda$ is related to the cosmological constant). For both spacetimes the parameter $\mu$ is related to the mass of the black holes. Here we study the region $r > r_H$ in Schwarzschild black hole, and the region $r_C > r > r_H$ in SdS black hole,  where $r_H$ denotes the radius of the event horizon and $r_C$ denotes the radius of the cosmological horizon.

\begin{figure}[t]
\begin{minipage}[b]{7.3cm}
\centering
\includegraphics[width=6.1cm]{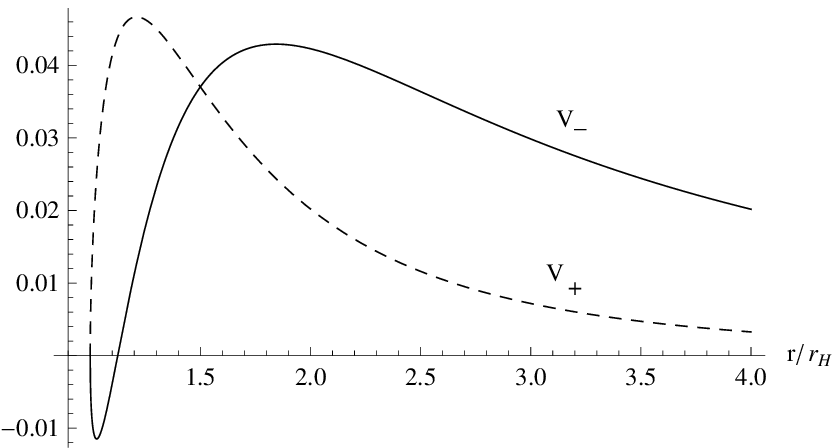}
\caption{Effective potentials $V_+$ (dashed) and $V_-$ (continuous) for the Schwarzschild black hole with $D=4$, $\mu=1$, and $l=0$.}
\label{fig: Schwarzschild vplus vminus}
\end{minipage}
\hspace{0.3cm}
\begin{minipage}[b]{7.3cm}
\centering
\includegraphics[width=6.1cm]{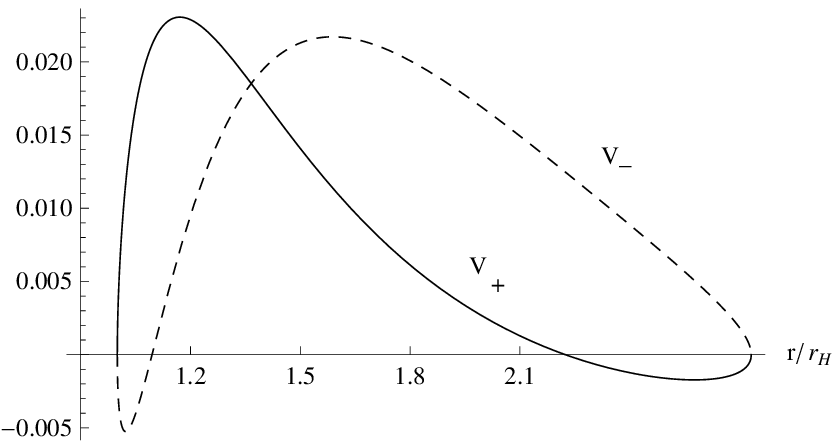}
\caption{Effective potentials $V_+$ (continuous) and $V_-$ (dashed) for the SdS black hole with $D=4$, $\mu=1$, and $l=0$.}
\label{fig: SdS vplus vminus}
\end{minipage}
\end{figure}

Thus in both black holes we obtain that the function $W$ of the formula (\ref{eq: W function general}) is equal to
\begin{equation} \label{eq: W function Schwarzschild SdS}
 W = \left( l + \frac{D-2}{2} \right) \frac{\sqrt{f}}{r},
\end{equation} 
(see the formulas (\ref{eq: metric function Schwarzschild}) and (\ref{eq: metric function Schwarzschild de Sitter})) and therefore the effective potentials (\ref{eq: potentials plus minus}) are equal to
\begin{equation} \label{eq: effective potentials general}
 V_\pm = \left( l + \frac{D-2}{2} \right)^2 \frac{f}{r^2} \pm \left( l + \frac{D-2}{2} \right) \sqrt{f} \left( \frac{1}{2r} \frac{\dd f}{\dd r} - \frac{f}{r^2} \right).
\end{equation} 

Taking into account that for the $D$-dimensional Schwarzschild (\ref{eq: metric function Schwarzschild}) and SdS (\ref{eq: metric function Schwarzschild de Sitter}) black holes their metric functions satisfy 
\begin{equation}
 \frac{1}{2r} \frac{\dd f}{\dd r} - \frac{f}{r^2}  = \frac{\mu (D-1)}{r^{D-1}} - \frac{1}{r^2},
\end{equation} 
we find that in these two spacetimes the effective potentials (\ref{eq: effective potentials general}) transform into
\begin{equation} \label{eq: effective potentials general two}
 V_\pm = \left( l + \frac{D-2}{2} \right) \left[ \left( l + \frac{D-2}{2} \right)  \frac{f}{r^2} \pm \frac{\sqrt{f}}{r^2}\left(\frac{\mu (D-1)}{r^{D-3}}  - 1 \right) \right] .
\end{equation} 

In Figs.\ \ref{fig: Schwarzschild vplus vminus} and \ref{fig: SdS vplus vminus}, for allowed values of the physical quantities $\mu$, $\lambda$, and $D$, we plot the effective potentials $V_+$ and $V_-$ in Schwarzschild and SdS black holes. In Fig.\ \ref{fig: Schwarzschild vplus vminus} we see that in Schwarzschild spacetime the effective potential $V_-$ is not positive definite, whereas in Fig.\ \ref{fig: SdS vplus vminus} we observe that in SdS background both effective potentials $V_+$ and $V_-$ are not positive definite. In the following we show that these behaviors of $V_+$ and $V_-$ are generic in $D$-dimensional Schwarzschild and SdS spacetimes.

\begin{figure}[t]
\begin{minipage}[b]{7.3cm}
\centering
\includegraphics[width=6.1cm]{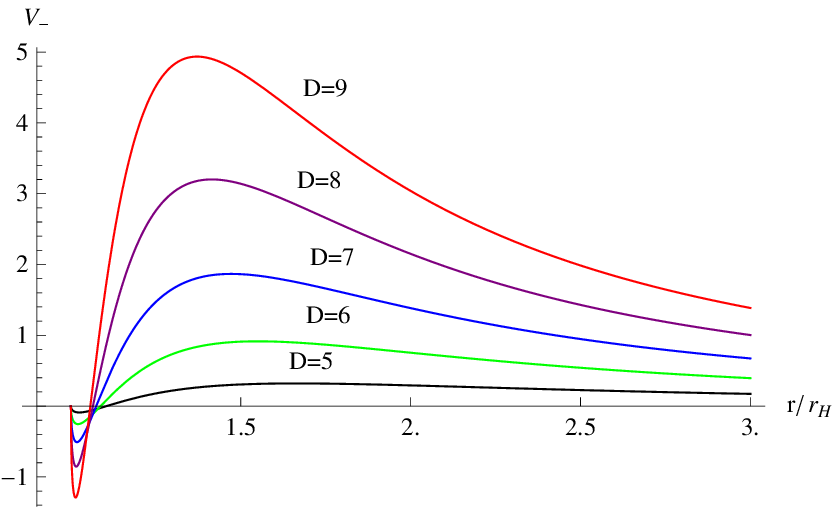}
\caption{Effective potentials $V_-$ for the Schwarzschild black hole with $\mu=1$, $l=0$, and $D=5,6,7,8,9$.}
\label{fig: Schwarzschild vminus dimension}
\end{minipage}
\hspace{0.3cm}
\begin{minipage}[b]{7.3cm}
\centering
\includegraphics[width=6.1cm]{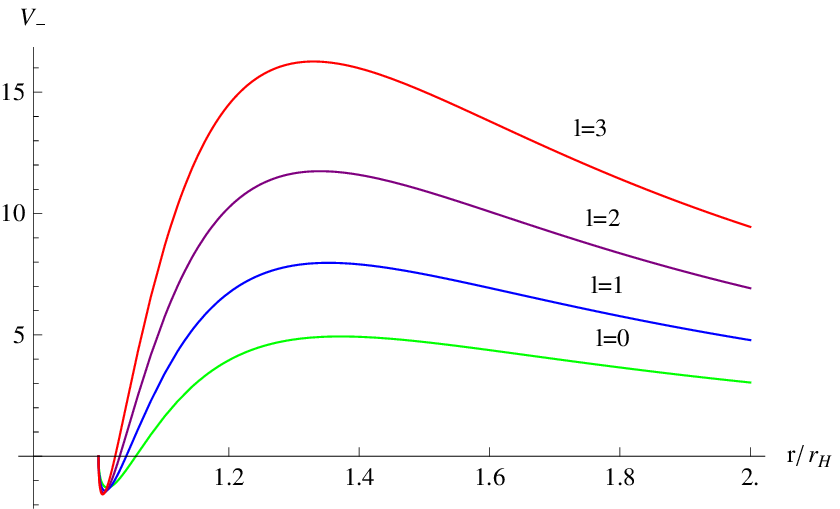}
\caption{Effective potentials $V_-$ for the Schwarzschild black hole with $D=9$, $\mu=1$, and $l=0,1,2,3$.}
\label{fig: Schwarzschild vminus angular}
\end{minipage}
\end{figure}

\subsection{Schwarzschild black hole}

In $D$-dimensional Schwarzschild black hole, the effective potentials (\ref{eq: effective potentials general two}) take the form 
\begin{align} \label{eq: effective potentials Schwarzschild}
 V_\pm &=  \left( l + \frac{D-2}{2} \right)^2 \left(1- \frac{2 \mu}{r^{D-3}}\right) \frac{1}{r^2} \\ 
&\pm  \left( l + \frac{D-2}{2} \right)  \left(1- \frac{2 \mu}{r^{D-3}}\right)^{1/2} \left(\frac{\mu (D-1)}{r^{D-1}}  - \frac{1}{r^2} \right) . \nonumber
\end{align} 
From these expressions, in an analytic way we show the following:
\begin{enumerate}
 \item At the event horizon $r=r_H$ both effective potentials are equal to zero.
\item Near the event horizon, that is, at $r=r_H + \delta$, with $\delta > 0$, $\delta \ll r_H$, we find that $V_+$ and $V_-$ fulfill
\begin{equation}
 V_+(r_H + \delta) > 0, \qquad \qquad  V_-(r_H + \delta) < 0.
\end{equation} 
\item If $r_0$ is the positive root of 
\begin{equation} \label{eq: equation for r0}
 \frac{\mu (D-1)}{r^{D-3}} - 1 = 0,
\end{equation} 
then
\begin{equation}
 V_+(r_0) = V_-(r_0) = \left( l + \frac{D-2}{2} \right)^2 \frac{f(r_0)}{r_0^2} > 0, 
\end{equation} 
that is, the effective potentials $V_+$ and $V_-$ take the same value at $r_0$. Notice that for $D \geq 4$ we find  that $r_0 > r_H$.
\item As $r \to \infty$, $V_+$ and $V_-$ go to zero taking positive values.
\end{enumerate}

\begin{figure}[t]
\begin{minipage}[b]{7.3cm}
\centering
\includegraphics[width=6.1cm]{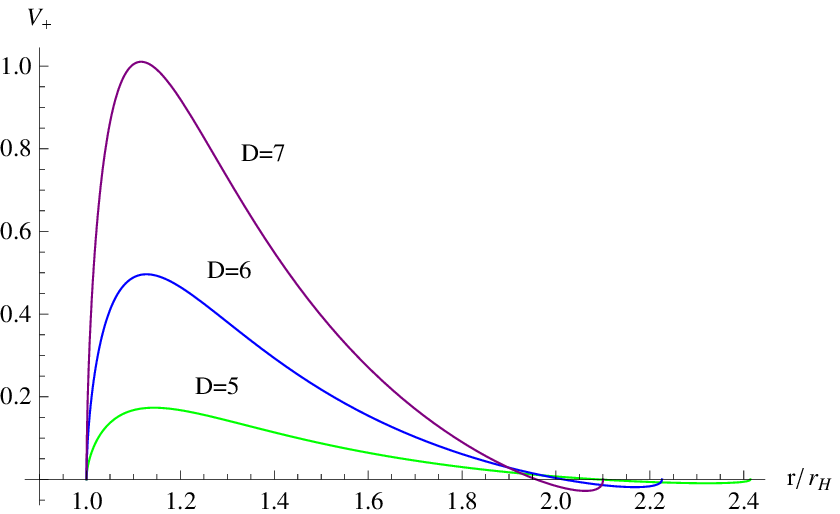}
\caption{Effective potentials $V_+$ for the SdS black hole with $\mu=1$, $l=0$, and  $D=5,6,7$.}
\label{fig: SdS vplus dimension}
\end{minipage}
\hspace{0.3cm}
\begin{minipage}[b]{7.3cm}
\centering
\includegraphics[width=6.1cm]{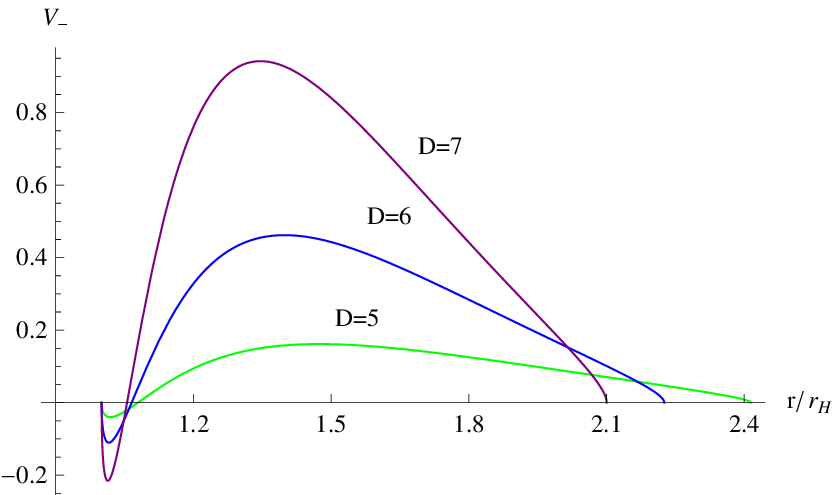}
\caption{Effective potentials $V_-$ for the SdS black hole with $\mu=1$, $l=0$, and  $D=5,6,7$.}
\label{fig: SdS vminus dimension}
\end{minipage}
\end{figure}

See Figs.\ \ref{fig: Schwarzschild vplus vminus}, \ref{fig: Schwarzschild vminus dimension}, and \ref{fig: Schwarzschild vminus angular}. Thus we get that in $D$-dimensional Schwarzschild black hole the effective potential $V_+$ is positive definite. Furthermore notice that the effective potentials $V_-$ have a small negative ditch near the event horizon  and therefore these are not positive definite (see Figs.\ \ref{fig: Schwarzschild vminus dimension} and \ref{fig: Schwarzschild vminus angular} for plots of the effective potentials $V_-$ for different values of the quantities $D$ and $l$).

\subsection{SdS black hole}

In $D$-dimensional SdS black hole the effective potentials (\ref{eq: effective potentials general two}) are equal to
\begin{align} \label{eq: effective potentials SdS}
 V_\pm &=   \left( l + \frac{D-2}{2} \right)^2 \left(1- \frac{2 \mu}{r^{D-3}} - \lambda r^2 \right) \frac{1}{r^2}  \\
&\pm  \left( l + \frac{D-2}{2} \right) \left(1- \frac{2 \mu}{r^{D-3}} - \lambda r^2 \right)^{1/2} \left(\frac{\mu (D-1)}{r^{D-1}}  - \frac{1}{r^2} \right) . \nonumber
\end{align} 
For  these effective potentials we show the following:
\begin{enumerate}
 \item At the event horizon and at the cosmological horizon both effective potentials are equal to zero.
\item Near the event horizon, that is, at $r=r_H + \delta$ we get that the effective potentials satisfy
 \begin{equation}
 V_+(r_H + \delta) > 0, \qquad \qquad  V_-(r_H + \delta) < 0,
\end{equation} 
and near the cosmological horizon, that is, at $r=r_C-\delta$ we find
\begin{equation}
 V_+(r_C - \delta) < 0, \qquad \qquad  V_-(r_C - \delta) > 0.
\end{equation} 
\item As for Schwarzschild spacetime, if $r_0$ denotes the positive root of Eq.\ (\ref{eq: equation for r0}), then in the $D$-dimensional non-extreme SdS black hole the radii $r_H$, $r_0$, $r_C$ fulfill $r_H < r_0 < r_C$, and the effective potentials take the same value at $r_0$, that is $V_+(r_0)=V_-(r_0)$. Notice that in SdS background the quantity $r_0$ corresponds to the common value for the radii of the event horizon and cosmological horizon in the extremal limit.
\end{enumerate}

\begin{figure}[t]
\begin{minipage}[b]{7.3cm}
\centering
\includegraphics[width=6.0cm]{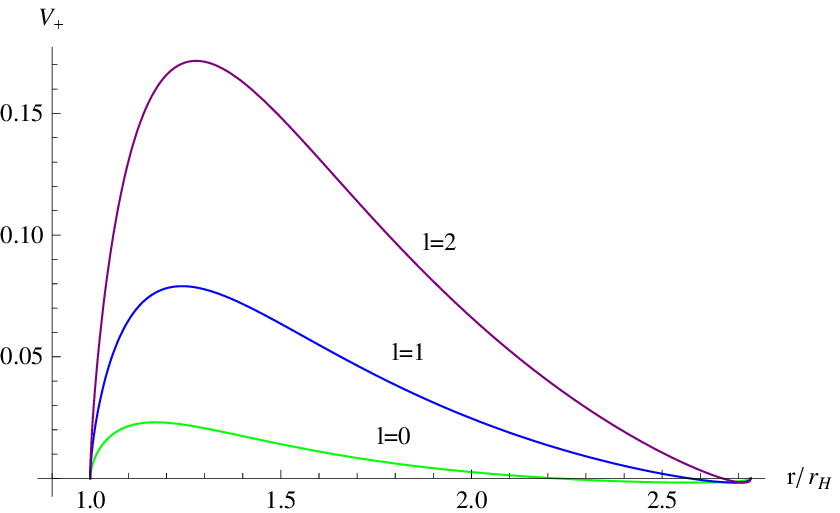}
\caption{Effective potentials $V_+$ for the SdS black hole with $\mu=1$, $D=4$, and  $l=0,1,2$.}
\label{fig: SdS vplus angular}
\end{minipage}
\hspace{0.3cm}
\begin{minipage}[b]{7.3cm}
\centering
\includegraphics[width=6.0cm]{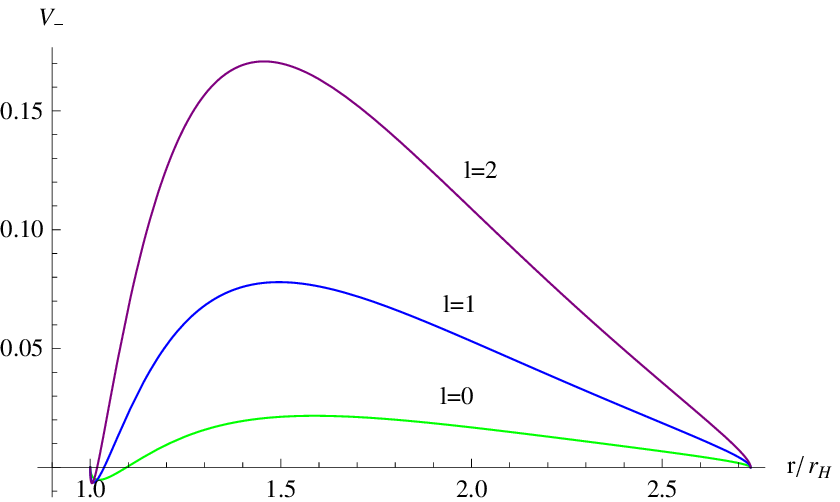}
\caption{Effective potentials $V_-$ for the SdS black hole with $\mu=1$, $D=4$, and  $l=0,1,2$.}
\label{fig: SdS vminus angular}
\end{minipage}
\end{figure}

Hence in $D$-dimensional SdS black hole, neither $V_+$ nor $V_-$ are positive definite, both effective potentials are negative in some interval, $V_+$ near the cosmological horizon and $V_-$ near the event horizon, (see Figs.\ \ref{fig: SdS vplus vminus}, \ref{fig: SdS vplus dimension}--\ref{fig: SdS vminus angular} for plots of these two effective potentials for different values of the parameters $D$ and $l$).

\section{Stability analysis}
\label{sec: Stability analysis}

For the $D$-dimensional Schwarzschild and SdS black holes the region exterior to the event horizon is static, therefore in this region Eqs.\ (\ref{eq: wavelike equations Dirac}) simplify to eigenvalue problems of the type
\begin{equation}
 \omega^2 R_\pm = A_\pm R_\pm ,
\end{equation} 
where $A_\pm$ are the operators
\begin{equation} \label{eq: A operators}
 A_\pm = -\frac{\dd^2 }{\dd x^2} + V_\pm .
\end{equation} 

In the domain $C_0^\infty(x)$ of smooth functions with compact support in $x$, the operators $A_\pm$ are symmetric. To state the classical stability of the Schwarzschild and SdS black holes (or of other black hole) against massless Dirac perturbations we must show that the operators $A_\pm$ can be extended to positive self-adjoint operators in $C_0^\infty(x)$ \cite{Ishibashi:2003ap}--\cite{Ishibashi:2011ws}. As is known, for well behaved initial data the positivity of the self-adjoint operators $A_\pm$ ensures that the solutions of Eqs.\ (\ref{eq: wavelike equations Dirac}) remain bounded.

Since in $D$-dimensional Schwarzschild and SdS black holes at least one of the effective potentials $V_\pm$ is not positive definite, we can not guarantee the classical stability of these spacetimes against massless Dirac fields. Although we expect that these perturbations do not produce instabilities, it is certainly desirable to prove the classical stability of these black holes under this fermion field. Furthermore, another objective in the study of these two backgrounds is to expound in a simple setting the procedure that we use to prove the classical stability when the effective potentials of the massless Dirac field are not positive definite.

To show that the operators $A_\pm$ of the formula (\ref{eq: A operators}) can be extended to positive self-adjoint operators we use the $S$-deformation method \cite{Ishibashi:2003ap}--\cite{Ishibashi:2011ws}. Thus following Refs.\ \cite{Ishibashi:2003ap}--\cite{Ishibashi:2011ws} we define the differential operator
\begin{equation} \label{eq: differential operator definition}
 \tilde{D} = \frac{\dd }{\dd x} + S ,
\end{equation} 
where $S$ is a regular function of $x$. Using an integration by parts and taking into account that for $\Phi \in C_0^\infty(x)$ we can cancel the boundary terms to obtain \cite{Ishibashi:2003ap}--\cite{Ishibashi:2011ws}
\begin{equation}
 \int_{-\infty}^\infty \dd x \,\,\Phi^* \left( -\frac{\dd^2}{\dd x^2} + V \right) \Phi = \int_{-\infty}^\infty \dd x \,\, \left[ (\tilde{D} \Phi)^* (\tilde{D} \Phi) + \tilde{V} |\Phi|^2 \right],
\end{equation} 
where the new potential is equal to
\begin{equation} \label{eq: new potential definition}
 \tilde{V} = V + \frac{\dd S}{\dd x} - S^2 .
\end{equation} 

Thus if we find a function $S$ such that the new potential (\ref{eq: new potential definition}) satisfies $\tilde{V} \geq 0$, then we show the classical stability of the spacetime against the perturbation field \cite{Ishibashi:2003ap}--\cite{Ishibashi:2011ws}. Therefore in the examples for which the effective potentials are not positive definite, to prove the classical stability of the $D$-dimensional Schwarzschild and SdS black holes under massless Dirac perturbations,  we need to find the appropriate $S$ functions to obtain new effective potentials that fulfill $\tilde{V} \geq 0$.

\subsection{Schwarzschild black hole}

In Sect.\ \ref{sec: Effective potentials in} we see that in $D$-dimensional Schwarzschild black hole only the effective potential $V_-$ of the formula (\ref{eq: effective potentials Schwarzschild}) is not positive definite. Choosing
\begin{equation} \label{eq: S function minus Schwazschild}
 S_-= \left( l + \frac{D-2}{2} \right) \left(1- \frac{2 \mu}{r^{D-3}}\right)^{1/2} \frac{1}{r},
\end{equation} 
we obtain that
\begin{equation} \label{eq: new potential minus Schwarzschild}
 \tilde{V}_- = V_- + \frac{\dd S_-}{\dd x} - S^2_- = 0,
\end{equation} 
and therefore we see that the new potential satisfies $\tilde{V}_- \geq 0$. Since in $D$-dimensional Schwarzschild black hole we find that $V_+ \geq 0$ and $\tilde{V}_- \geq 0$  we state the classical stability of this black hole against massless Dirac perturbations.

\subsection{SdS black hole}

For the $D$-dimensional SdS black hole in Sect.\ \ref{sec: Effective potentials in} we observe that both potentials $V_\pm$ are not positive definite. First we focus on the effective potential $V_+$. For this we choose
\begin{equation} \label{eq: S function plus SdS}
 S_+= - \left( l + \frac{D-2}{2} \right) \left(1- \frac{2 \mu}{r^{D-3} } - \lambda r^2 \right)^{1/2}   \frac{1}{r},
\end{equation} 
to get
\begin{equation} \label{eq: new potential plus SdS}
 \tilde{V}_+ = V_+ + \frac{\dd S_+}{\dd x} - S^2_+ = 0 .
\end{equation} 
Hence, in the SdS spacetime we get that $\tilde{V}_+ \geq 0$. 

For the effective potential $V_-$ we choose (compare with the expression (\ref{eq: S function minus Schwazschild}) for the Schwarzschild black hole)
\begin{equation} \label{eq: S function minus SdS}
 S_-= \left( l + \frac{D-2}{2} \right) \left(1- \frac{2 \mu}{r^{D-3}} - \lambda r^2 \right)^{1/2} \frac{1}{r},
\end{equation} 
to find that in the SdS black hole the new potential (\ref{eq: new potential minus Schwarzschild}) satisfies $\tilde{V}_- \geq 0$. 

Thus with the help of the $S$-deformation method in the $D$-dimensional SdS spacetime  we obtain two new potentials $\tilde{V}_+$ and $\tilde{V}_-$ that satisfy $\tilde{V}_+ \geq 0$ and $\tilde{V}_- \geq 0$. Hence we state the classical stability of this black hole against massless Dirac fields.

It is convenient to note that the effective potentials (\ref{eq: effective potentials general two}) for $D=4$ are equal to the effective potentials given by Khanal and Panchapakesan \cite{Khanal:1981bb} for the massless Dirac field moving in the four-dimensional SdS black hole (see the formulas before Eq.\ (3.16) in Ref.\ \cite{Khanal:1981bb}). In the previous reference we do not find plots of the effective potentials for the massless Dirac field and in that reference is not noted that these effective potentials in the four-dimensional SdS black hole are negative in some intervals (see Figs.\ \ref{fig: SdS vplus vminus}, \ref{fig: SdS vplus angular}, \ref{fig: SdS vminus angular}), it is only noted that the effective potentials are of short range, that is, they go to zero at the black hole and cosmological horizons. Furthermore, issues related to the stability of the massless Dirac field propagating in four-dimensional SdS black hole are not analyzed in Ref.\ \cite{Khanal:1981bb}.

Notice that for the Dirac field propagating in Schwarzschild and SdS black holes the quantities $S_+$ and $S_-$ of the formulas (\ref{eq: S function minus Schwazschild}), (\ref{eq: S function plus SdS}), and (\ref{eq: S function minus SdS}) depend on the spacetime dimension $D$ and the nonnegative integer $l$. Hence given a spacetime and its dimension, due to the parameter $l$ for different field modes we get different expressions for the functions $S_+$ and $S_-$. Furthermore in these black holes the quantities $S_+$ and $S_-$ are proportional to the square root of the function $f$ defined in the formula (\ref{eq: identifications metric functions}).

\subsection{Generalization}

From the expressions (\ref{eq: S function minus Schwazschild}), (\ref{eq: S function plus SdS}), and (\ref{eq: S function minus SdS}) for the functions $S_-$ and $S_+$ in the Schwarzschild and SdS black holes we notice that in both spacetimes these functions fulfill 
\begin{equation}
 S_\pm \propto W,
\end{equation} 
where the function $W$ appears in the formula (\ref{eq: W function Schwarzschild SdS}).

Thus from the previous observation we get the following result. In a spacetime such that the equations of motion for a perturbing field simplify to a wavelike equation with an effective potential of the form 
\begin{equation}
 V_+ = W^2 + \frac{\dd W}{\dd x}, \qquad \left( V_- = W^2  - \frac{\dd W}{\dd x} \right),
\end{equation}  
if this effective potential is not positive definite, then using the $S$-deformation method and choosing
\begin{equation}  \label{eq: S functions general}
 S_+ = - W , \qquad (S_-= W) ,
\end{equation} 
we find that the new potential $\tilde{V}_+$ ($\tilde{V}_-$) satisfies $\tilde{V}_+ = 0$ ($\tilde{V}_- = 0$).

Therefore for a black hole with line element of the form (\ref{eq: maximally symmetric metric}) and for which the equations of motion for a perturbing field simplify to wavelike equations with effective potentials of the type (\ref{eq: potentials plus minus}), if we obtain that a potential is not positive definite, then using the $S$-deformation method \cite{Ishibashi:2003ap}--\cite{Ishibashi:2011ws}, with the appropriate $S$ function of the formulas (\ref{eq: S functions general}), we always find a new potential that is nonnegative, and hence we establish the classical stability of the black hole against this perturbing field.

Something similar happens in the static patch of the $D$-dimensional de Sitter spacetime when we show the stability of the quasinormal modes for the massless Dirac field \cite{LopezOrtega:2012vi}. 

Notice that the previous results are valid also for black holes with line element (\ref{eq: maximally symmetric metric}) whose metric functions $F$ and $G$ do not satisfy $F^2 = 1/G^2$. For this instance, in a straightforward way, we check that our method works  because in our calculations the metric functions $F$ and $G$ are involved as a part of the function $W$ defined in the formula (\ref{eq: W function general}) and of the tortoise coordinate $x$ given in the expression (\ref{eq: definitions Z and tortoise}). 

As an application of the previous results, we notice that in $D$-dimensional Reissner-Nordstr\"om and Reissner-Nordstr\"om de Sitter black holes the effective potentials for the massless Dirac field take the form (\ref{eq: potentials plus minus}). In the Reissner-Nordstr\"om spacetime the effective potentials behave similarly to those of the Schwarzschild spacetime, that is, $V_-$ is not positive definite. Moreover in Reissner-Nordstr\"om de Sitter black hole the effective potentials $V_\pm$ behave in a similar way to those of the SdS background, thus, neither $V_+$ nor $V_-$ are positive definite. Since in these two black holes the effective potentials are of the form (\ref{eq: potentials plus minus}) we can use our results to establish the stability of these two spacetimes against massless Dirac perturbations. 

For the $D$-dimensional Reissner-Nordstr\"om de Sitter spacetime our result about its classical stability with respect to massless Dirac perturbations is different from that of Ref.\ \cite{Konoplya:2008au} (see also Ref.\ \cite{Cardoso:2010rz}) since for $D \geq 7$ this black hole is unstable under the coupled electromagnetic and gravitational perturbations of scalar type \cite{Konoplya:2008au,Cardoso:2010rz}.

\section{Discussion}
\label{sec: Discussion}

Based on the method that we use to show the classical stability of the $D$-dimensional Schwarzschild and SdS black holes against massless Dirac perturbations, we are able to extend this procedure and prove the classical stability under massless Dirac perturbations of the maximally symmetric black holes (\ref{eq: maximally symmetric metric}), since in these spacetimes we reduce the massless Dirac equation to wavelike equations with effective potentials of the form (\ref{eq: potentials plus minus}), and if these are not positive definite, then using the $S$-deformation method we find new potentials that are nonnegative.

We notice that the method of Sect.\ \ref{sec: Stability analysis} finds another use. It is helpful to show the classical stability against other fields, as the gravitational (already proven in Refs.\ \cite{Ishibashi:2003ap}, \cite{Kodama:2003kk}), the electromagnetic and the Klein-Gordon perturbations (see below). For example, in General Relativity and for $D$-dimensional uncharged static spacetimes of the form (\ref{eq: maximally symmetric metric}) with $F^2 = 1/G^2=f$, the equations of motion for the vector type gravitational perturbations reduce to a wavelike equation (\ref{eq: wavelike equations Dirac}) with an effective potential equal to \cite{Ishibashi:2003ap} (see Eq.\ (2.17) of Ref.\ \cite{Ishibashi:2003ap})
\begin{equation} \label{eq: gravitational perturbations vector}
 V_V = \frac{f}{r^2}\left[ \kappa_V^2 - (D-3)K + \frac{(D-2)D}{4} f - \frac{D-2}{2} r \frac{\dd f}{\dd r} \right],
\end{equation} 
where $\kappa_V^2$ are the eigenvalues of the vector harmonics on the base manifold with metric $\dd \Sigma_{D-2}^2$ and $K$ is a discrete parameter related to the scalar curvature of the base manifold \cite{Ishibashi:2003ap}--\cite{Ishibashi:2011ws}.

In a similar way, for $D \geq 5$ the effective potential for the tensor type gravitational perturbations is equal to \cite{Kodama:2003kk} (see Eq.\ (3.7) of Ref.\ \cite{Kodama:2003kk})
\begin{equation} \label{eq: gravitational perturbations tensor}
 V_T = \frac{f}{r^2}\left[ \kappa_L - 2(D-3)K + \frac{(D-2)(D-4)}{4} f + \frac{D-2}{2} r \frac{\dd f}{\dd r} \right],
\end{equation} 
where $\kappa_L $ are the eigenvalues of the Lichnerowicz operator on the base manifold. Kodama and Ishibashi show that for some spherically symmetric $D$-dimensional black holes the effective potentials (\ref{eq: gravitational perturbations vector}) and (\ref{eq: gravitational perturbations tensor}) are not positive definite \cite{Ishibashi:2003ap,Kodama:2003kk}.

Defining the function
\begin{equation} \label{eq: W for gravitational perturbations}
 W_{GP} = \frac{D-2}{2} \frac{f}{r},
\end{equation} 
we find that the effective potentials $V_V$ and $V_T$ take the form 
\begin{align} \label{eq: potentials GP W}
 V_V &= (\kappa_V^2 - (D-3)K) \frac{f}{r^2} + W_{GP}^2 - f\frac{\dd  W_{GP}}{\dd r} , \\
V_T &=  (\kappa_L - 2(D-3)K) \frac{f}{r^2} + W_{GP}^2 + f\frac{\dd  W_{GP}}{\dd r}. \nonumber 
\end{align}

Thus except for the first term, the mathematical form of these effective potentials is similar to that of the $V_\pm$ given in the formula (\ref{eq: potentials plus minus}) for the massless Dirac field. Hence using the $S$-deformation method with the functions
\begin{equation} \label{eq: S functions GP}
 S_V = W_{GP} , \qquad S_T = - W_{GP} , 
\end{equation}  
we obtain the new potentials 
\begin{align} \label{eq: new potentials GP}
 \tilde{V}_V &= V_V + f\frac{\dd S_V}{\dd r} - S_V^2 = (\kappa_V^2 - (D-3)K) \frac{f}{r^2} , \\
\tilde{V}_T &= V_T + f\frac{\dd S_T}{\dd r} - S_T^2 =  (\kappa_L - 2(D-3)K) \frac{f}{r^2}. \nonumber
\end{align} 

With the help of the $S$-deformation method in Refs.\ \cite{Ishibashi:2003ap}, \cite{Kodama:2003kk}, Kodama and Ishibashi obtain the new potentials $\tilde{V}_V$ and $\tilde{V}_T$ of the formulas (\ref{eq: new potentials GP}) (see the expressions (2.22) of Ref.\ \cite{Ishibashi:2003ap} and (6.8) of Ref.\ \cite{Kodama:2003kk}), but they find the functions $S_V$ and $S_T$ ``\textit{by inspection}'' (see the formulas (2.21) of Ref.\ \cite{Ishibashi:2003ap} and (6.7) of Ref.\ \cite{Kodama:2003kk}).\footnote{Since we find the same potentials $\tilde{V}_V$ and $\tilde{V}_T$ of Refs.\ \cite{Ishibashi:2003ap}, \cite{Kodama:2003kk}, \cite{Ishibashi:2011ws} we obtain the same conclusions  about the classical stability of these black holes with respect to the gravitational perturbations of vector and tensor type.} Here motivated by our results for the massless Dirac field, we expound a more systematic procedure to find the functions $S_V$ and $S_T$ of the formulas (\ref{eq: S functions GP}). 

Thus taking into account our previous results, for a black hole such that the equations of motion for a perturbation simplify to wavelike equations with effective potentials of the type
\begin{equation}
 V_\pm = E_\pm + W^2 \pm \frac{\dd  W}{\dd x} , \\
\end{equation}
we find that the functions $E_\pm$ determine its classical stability, as already shown in Refs.\ \cite{Ishibashi:2003ap}--\cite{Ishibashi:2011ws} for the gravitational perturbations of vector and tensor type. 

For the four-dimensional Schwarzschild black hole, for which $F^2 = 1/G^2 = f = 1 - 2\mu/r $, $H(r)^2=r^2$, and $\dd \Sigma_{D-2}^2$ is the line element of the unit 2-sphere, it is known that the effective potentials  for the axial (vector) and polar (scalar) type gravitational perturbations are \cite{Chandrasekhar book,Chandrasekhar:1975zz} 
\begin{eqnarray} \label{eq: potentials axial polar}
V_a &=& \frac{f}{r^3}\left[ k(k+1)- 6\mu \right], \nonumber \\ 
V_p & =& \frac{2 f}{r^3 (n r + 3 \mu)^2 } \left[ n^2 (n+1) r^3 + 3 \mu n^2 r^2 + 9 \mu^2 n r + 9 \mu^3 \right],
\end{eqnarray}
where $k$ denotes the azimuthal number ($k= 2, 3,4, \dots$, for the gravitational perturbations) and $n=(k^2 + k -2)/2$.

In four-dimensional Schwarzschild black hole the effective potentials (\ref{eq: potentials axial polar}) are positive definite, hence the four-dimensional Schwarzschild black hole is stable against gravitational perturbations \cite{Wald:1979,Ishibashi:2003ap}. Furthermore the effective potentials for the axial and polar perturbations take the form \cite{Chandrasekhar book,Chandrasekhar:1975zz}
\begin{equation} \label{eq: schwarzschild axial polar susy form}
 V_{p,a} = \pm \beta \frac{\dd P}{\dd x} + \beta^2 P^2 + \varepsilon P,
\end{equation} 
where the upper sign corresponds to polar perturbations, the lower sign to axial perturbations and
\begin{align}
 \beta = 6 \mu , \quad \quad
\varepsilon  = (k-1)k(k+1)(k+2), \quad \quad
P  = \frac{f}{r} \frac{1}{(k-1)(k+2) r + 6 \mu}. 
\end{align}

Therefore from our previous results we find that the stability of the axial and polar perturbations is determined by the factor $\varepsilon P$  of the formulas (\ref{eq: schwarzschild axial polar susy form}) when we choose $W = \mp \beta P$, since we get the new potentials
\begin{equation}
 \tilde{V}_{p,a} = \varepsilon P = (k-1)k(k+1)(k+2) P,
\end{equation}  
and since outside the event horizon $\tilde{V} \geq 0$, we find the already known result that the four-dimensional Schwarzschild black hole is stable against axial and polar perturbations \cite{Wald:1979,Ishibashi:2003ap}.

Similar considerations apply to the tensor type gravitational perturbation of the spherically symmetric Einstein-Gauss-Bonnet black holes, because its effective potential takes a mathematical form similar to that of $V_V$ and $V_T$ in the expressions (\ref{eq: potentials GP W}) (see the formulas (16) and (18) of Ref.\ \cite{Dotti:2004sh}). Thus in this example we can use our previous result to find the appropriate $S$ function and study the classical stability against gravitational perturbations of tensor type \cite{Dotti:2004sh,Dotti:2005sq}.

For the maximally symmetric black holes (\ref{eq: maximally symmetric metric}) when the base manifold $\dd \Sigma^2_{D-2}$ is a $(D-2)$-dimensional sphere and the function $H$ satisfies $H(r)^2 = r^2$, the equations of motion for the Klein-Gordon and electromagnetic fields simplify to Schr\"odinger type equations with effective potentials equal to
\begin{equation} \label{eq: effective potential Klein-Gordon}
 V_{KG} = \frac{k(k+D-3) F^2}{r^2} + m^2 F^2 + \frac{(D-2)(D-4)}{4 r^2} \frac{F^2}{G^2} + \frac{D-2}{2r}\frac{F}{G} \left( \frac{\dd }{\dd r} \frac{F}{G} \right),
\end{equation}  
for the Klein-Gordon field and
\begin{align} \label{eq: effective potentials electromagnetic}
 V_I &=  \frac{k(k+D-3) F^2}{r^2} + \frac{(D-2)(D-4)F^2}{4 r^2 G^2} - \frac{(D-4)}{4 r} \left( \frac{\dd }{\dd r } \frac{F^2}{G^2} \right) ,\nonumber \\
V_{II} &=  \frac{(k+1)(k+D-4) F^2}{r^2} + \frac{(D-4)(D-6)F^2}{4 r^2 G^2} + \frac{(D-4)}{4 r} \left( \frac{\dd }{\dd r } \frac{F^2}{G^2} \right),
\end{align}
for the modes I and II of the electromagnetic field \cite{Crispino:2000jx}. In the formula (\ref{eq: effective potential Klein-Gordon}) and in what follows, $m$ denotes the mass of the Klein-Gordon field. As previously, $k$ denotes the azimuthal number, but $k=0,1,2,\dots$, for the Klein-Gordon field and $k=1,2,3,\dots$, for the electromagnetic field.

We notice that the effective potential for the Klein-Gordon field (\ref{eq: effective potential Klein-Gordon}) is not positive definite in $D$-dimensional  Reissner-N\"ordstrom de Sitter black hole. Furthermore for the electromagnetic field the effective potential $V_I$ is not positive definite in $D$-dimensional SdS black hole. Although we do not know a maximally symmetric spacetime for which the effective potential $V_{II}$ is negative in some interval, for example, $V_{II}$ is nonnegative in Schwarzschild and SdS black holes, without any problem we include this effective potential in the discussion that follows.

Making some algebraic operations the effective potentials $V_{KG}$, $V_I$, and $V_{II}$ take the form 
\begin{align} \label{eq: KleinGordon electromagnetic susy}
 V_{KG} & = \left( \frac{k(k+D-3) }{r^2} + m^2\right)  F^2 + W_{KG}^2 + \frac{\dd W_{KG}}{\dd x} , \nonumber \\
V_I & = \frac{k(k+D-3) F^2}{r^2} + W_{EM}^2 - \frac{\dd W_{EM}}{\dd x}, \\
V_{II} &=  \frac{(k+1)(k+D-4) F^2}{r^2} + W_{EM}^2 + \frac{\dd W_{EM}}{\dd x}, \nonumber 
\end{align}
where
\begin{equation}
 W_{KG} = \frac{D-2}{2r}\frac{F}{G}, \qquad \qquad W_{EM} = \frac{D-4}{2r}\frac{F}{G} .
\end{equation} 
For spherically symmetric spacetimes that satisfy $F^2 = 1/G^2=f$ and if $F \geq 0$ and $G \geq 0$ (as outside the event horizon in Schwarzschild and SdS black holes) we find 
\begin{equation}
 W_{KG} = \frac{D-2}{2}\frac{f}{r}, \qquad \qquad W_{EM} = \frac{D-4}{2}\frac{f}{r} .
\end{equation} 
Thus under these conditions we obtain $W_{KG} = W_{GP}$ (see the formula (\ref{eq: W for gravitational perturbations})).

Hence the effective potentials $V_{KG}$, $V_I$, and $V_{II}$ of the formulas (\ref{eq: KleinGordon electromagnetic susy}) take a similar mathematical form that the effective potentials $V_V$ and $V_T$ of the formulas (\ref{eq: potentials GP W}) for the gravitational perturbations. Using the $S$-deformation method with the functions
\begin{equation}
 S_{KG} = - W_{KG}, \qquad S_I = W_{EM}, \qquad S_{II} = -W_{EM} ,
\end{equation} 
for the Klein-Gordon and electromagnetic fields we obtain the new potentials
\begin{align} \label{eq: new effective potentials KG EM}
 \tilde{V}_{KG} & = \left( \frac{k(k+D-3) }{r^2} + m^2\right)  F^2 , \nonumber \\
 \tilde{V}_I & = \frac{k(k+D-3) F^2}{r^2} , \\
\tilde{V}_{II} &=  \frac{(k+1)(k+D-4) F^2}{r^2} . \nonumber 
\end{align}

The new effective potentials (\ref{eq: new effective potentials KG EM}) satisfy $\tilde{V}_{KG} \geq 0$, $\tilde{V}_I \geq 0$, and $\tilde{V}_{II} \geq 0$ outside the event horizon of the $D$-dimensional maximally symmetric black holes (\ref{eq: maximally symmetric metric}) with $(D-2)$-dimensional spheres as base manifolds. Hence we can assert that these black holes are stable against Klein-Gordon and electromagnetic perturbations.  

As for the massless Dirac field, for the gravitational, electromagnetic, and Klein-Gordon perturbations the functions $S_V$, $S_T$, $S_{KG}$, $S_I$, and $S_{II}$ depend on the spacetime dimension, but in contrast to the corresponding functions for the massless Dirac field, these functions are proportional to $f$ (defined in the formula (\ref{eq: identifications metric functions})) and they do not depend on the azimuthal number. Thus given the spacetime and its dimension, for all the modes of the field we find only one expression for each of these functions.  

We believe that deserves further research to analyze the usefulness of this method to state the classical stability of other black holes. Moreover it is appropriate to extend this work and prove the classical stability of the maximally symmetric black holes (\ref{eq: maximally symmetric metric}) against massive and charged Dirac fields.

\section{Acknowledgments}

This work was supported by CONACYT M\'exico, SNI M\'exico, EDI-IPN, COFAA-IPN, and Research Projects SIP-20120773 and SIP-20121648.


\begin{thebibliography}{}


\bibitem{Kokkotas:1999bd}
  K.~D.~Kokkotas and B.~G.~Schmidt,
  Living Rev.\ Rel.\  {\bf 2}, 2 (1999)
  [arXiv:gr-qc/9909058].

\bibitem{Nollert:1999bd}
H.~P.~Nollert, 
Class.\ Quantum Grav.\ {\bf 16}, R159 (1999).

\bibitem{Berti:2009kk}
  E.~Berti, V.~Cardoso and A.~O.~Starinets,
  Class.\ Quant.\ Grav.\  {\bf 26}, 163001 (2009)
  [arXiv:0905.2975 [gr-qc]].

\bibitem{Konoplya:2011qq}
  R.~A.~Konoplya and A.~Zhidenko,
  Rev.\ Mod.\ Phys.\  {\bf 83}, 793 (2011)
  [arXiv:1102.4014 [gr-qc]].

\bibitem{Regge:1957td}
  T.~Regge and J.~A.~Wheeler,
  Phys.\ Rev.\  {\bf 108}, 1063 (1957).

\bibitem{Zerilli:1974ai}
  F.~J.~Zerilli,
  Phys.\ Rev.\  D {\bf 9}, 860 (1974).

\bibitem{Zerilli:1971wd}
  F.~J.~Zerilli,
  Phys.\ Rev.\  D {\bf 2}, 2141 (1970).

\bibitem{Vishveshwara:1970zz}
  C.~V.~Vishveshwara,
  Nature {\bf 227}, 936 (1970).

\bibitem{Wald:1979}
  R.~M.~Wald,
J.\ Math.\ Phys.\ {\bf 20}, 1056 (1979).

\bibitem{Ishibashi:2003ap}
  A.~Ishibashi and H.~Kodama,
  Prog.\ Theor.\ Phys.\  {\bf 110}, 901 (2003)
  [arXiv:hep-th/0305185].

\bibitem{Kodama:2003kk} 
  H.~Kodama and A.~Ishibashi,
  Prog.\ Theor.\ Phys.\  {\bf 111}, 29 (2004)
  [hep-th/0308128].

\bibitem{Ishibashi:2011ws} 
  A.~Ishibashi and H.~Kodama,
  Prog.\ Theor.\ Phys.\ Suppl.\  {\bf 189}, 165 (2011)
  [arXiv:1103.6148 [hep-th]].

\bibitem{Gibbons:2002pq} 
  G.~Gibbons and S.~A.~Hartnoll,
  Phys.\ Rev.\ D {\bf 66}, 064024 (2002)
  [hep-th/0206202].

\bibitem{Dotti:2004sh} 
  G.~Dotti and R.~J.~Gleiser,
  Class.\ Quant.\ Grav.\  {\bf 22}, L1 (2005)
  [gr-qc/0409005].

\bibitem{Dotti:2005sq} 
  G.~Dotti and R.~J.~Gleiser,
  Phys.\ Rev.\ D {\bf 72}, 044018 (2005)
  [gr-qc/0503117].

\bibitem{Gleiser:2005ra} 
  R.~J.~Gleiser and G.~Dotti,
  Phys.\ Rev.\ D {\bf 72}, 124002 (2005)
  [gr-qc/0510069].

\bibitem{Beroiz:2007gp} 
  M.~Beroiz, G.~Dotti and R.~J.~Gleiser,
  Phys.\ Rev.\ D {\bf 76}, 024012 (2007)
  [hep-th/0703074].

\bibitem{Takahashi:2009dz} 
  T.~Takahashi and J.~Soda,
  Phys.\ Rev.\ D {\bf 79}, 104025 (2009)
  [arXiv:0902.2921 [gr-qc]].

\bibitem{Takahashi:2009xh} 
  T.~Takahashi and J.~Soda,
  Phys.\ Rev.\ D {\bf 80}, 104021 (2009)
  [arXiv:0907.0556 [gr-qc]].

\bibitem{Neupane:2003vz} 
  I.~P.~Neupane,
  Phys.\ Rev.\ D {\bf 69}, 084011 (2004)
  [hep-th/0302132].

\bibitem{Birmingham:2007yv} 
  D.~Birmingham and S.~Mokhtari,
  Phys.\ Rev.\ D {\bf 76}, 124039 (2007)
  [arXiv:0709.2388 [hep-th]].

\bibitem{Konoplya:2008au} 
  R.~A.~Konoplya and A.~Zhidenko,
  Phys.\ Rev.\ Lett.\  {\bf 103}, 161101 (2009)
  [arXiv:0809.2822 [hep-th]].

\bibitem{Cardoso:2010rz} 
  V.~Cardoso, M.~Lemos and M.~Marques,
  Phys.\ Rev.\ D {\bf 80}, 127502 (2009)
  [arXiv:1001.0019 [gr-qc]].

\bibitem{Emparan:2008eg} 
  R.~Emparan and H.~S.~Reall,
  Living Rev.\ Rel.\  {\bf 11}, 6 (2008)
  [arXiv:0801.3471 [hep-th]].

\bibitem{Martellini:1977qf}
  M.~Martellini and A.~Treves,
  Phys.\ Rev.\  D {\bf 15}, 3060 (1977);

\bibitem{Unruh:1973}
W.~G.~Unruh, 
Phys.\ Rev.\  Lett.\ {\bf 31}, 1265 (1973).

\bibitem{Iyer:1978du}
  B.~R.~Iyer and A.~Kumar,
  Phys.\ Rev.\  D {\bf 18}, 4799 (1978).

\bibitem{Gibbons:1993hg}
  G.~W.~Gibbons and A.~R.~Steif,
  Phys.\ Lett.\  B {\bf 314}, 13 (1993)
  [arXiv:gr-qc/9305018];

\bibitem{Das:1996we}
  S.~R.~Das, G.~W.~Gibbons and S.~D.~Mathur,
  Phys.\ Rev.\ Lett.\  {\bf 78}, 417 (1997)
  [arXiv:hep-th/9609052].

\bibitem{LopezOrtega:2004cq} 
  A.~Lopez-Ortega,
  Gen.\ Rel.\ Grav.\  {\bf 36}, 1299 (2004).
  
\bibitem{LopezOrtega:2009qc}
  A.~Lopez-Ortega,
  Lat.\ Am.\ J.\ Phys.\ Educ.\  {\bf 3}, 578 (2009)
  [arXiv:0906.2754 [gr-qc]].

\bibitem{Cho:2007zi}
  H.~T.~Cho, A.~S.~Cornell, J.~Doukas and W.~Naylor,
  Phys.\ Rev.\  D {\bf 75}, 104005 (2007)
  [arXiv:hep-th/0701193].
 
\bibitem{Cho:2007ce}
  H.~T.~Cho, A.~S.~Cornell, J.~Doukas and W.~Naylor,
  Phys.\ Rev.\  D {\bf 77}, 041502 (2008)
  [arXiv:0710.5267 [hep-th]].

\bibitem{Cho:2007de} 
  H.~T.~Cho, A.~S.~Cornell, J.~Doukas and W.~Naylor,
  Phys.\ Rev.\ D {\bf 77}, 016004 (2008)
  [arXiv:0709.1661 [hep-th]].

\bibitem{Rogatko:2009jp} 
  M.~Rogatko and A.~Szyplowska,
  Phys.\ Rev.\ D {\bf 79}, 104005 (2009)
  [arXiv:0904.4544 [hep-th]].

\bibitem{Chakrabarti:2008xz} 
  S.~K.~Chakrabarti,
  Eur.\ Phys.\ J.\ C {\bf 61}, 477 (2009)
  [arXiv:0809.1004 [gr-qc]].
  
\bibitem{Zhidenko:2008fp} 
  A.~Zhidenko,
  Phys.\ Rev.\ D {\bf 78}, 024007 (2008)
  [arXiv:0802.2262 [gr-qc]].

\bibitem{LopezOrtega:2007sr} 
  A.~Lopez-Ortega,
  Gen.\ Rel.\ Grav.\  {\bf 39}, 1011 (2007)
  [arXiv:0704.2468 [gr-qc]].

\bibitem{LopezOrtega:2009zx}
  A.~Lopez-Ortega,
  Int.\ J.\ Mod.\ Phys.\ D {\bf 9}, 1441 (2009)
  [arXiv:0905.0073 [gr-qc]].

\bibitem{LopezOrtega:2010uu} 
  A.~Lopez-Ortega,
  Rev.\ Mex.\ Fis.\  {\bf 56}, 44 (2010)
  [arXiv:1006.4906 [gr-qc]].

\bibitem{LopezOrtega:2010tg}
  A.~Lopez-Ortega,
  Class.\ Quant.\ Grav.\  {\bf 28}, 035009 (2011)
  [arXiv:1003.4248 [gr-qc]].

\bibitem{Kanti:2006ua} 
  P.~Kanti, R.~A.~Konoplya and A.~Zhidenko,
  Phys.\ Rev.\ D {\bf 74}, 064008 (2006)
  [gr-qc/0607048].

\bibitem{Kanti:2005xa} 
  P.~Kanti and R.~A.~Konoplya,
  Phys.\ Rev.\ D {\bf 73}, 044002 (2006)
  [hep-th/0512257].

\bibitem{Oikonomou:2012js} 
  V.~K.~Oikonomou,
  arXiv:1204.2395 [gr-qc].

\bibitem{Cotaescu:1998ay} 
  I.~I.~Cotaescu,
  Mod.\ Phys.\ Lett.\ A {\bf 13}, 2991 (1998)
  [gr-qc/9808030].

\bibitem{Cotaescu:2003be} 
  I.~I.~Cotaescu,
  Int.\ J.\ Mod.\ Phys.\ A {\bf 19}, 2217 (2004)
  [gr-qc/0306127].

\bibitem{Oota:2007vx} 
  T.~Oota and Y.~Yasui,
  Phys.\ Lett.\ B {\bf 659}, 688 (2008)
  [arXiv:0711.0078 [hep-th]].

\bibitem{Wu:2008df} 
  S.~Q.~Wu,
  Phys.\ Rev.\ D {\bf 78}, 064052 (2008)
  [arXiv:0807.2114 [hep-th]].

\bibitem{Wu:2008bc} 
  S.~Q.~Wu,
  Class.\ Quant.\ Grav.\  {\bf 26}, 055001 (2009)
  [Erratum-ibid.\  {\bf 26}, 189801 (2009)]
  [arXiv:0808.3435 [hep-th]].

\bibitem{Chandrasekhar book} S.~Chandrasekhar, {\it The Mathematical Theory of Black Holes}, (Oxford University Press, Oxford, 1983).

\bibitem{Gundlach:1993tp} 
  C.~Gundlach, R.~H.~Price and J.~Pullin,
  Phys.\ Rev.\ D {\bf 49}, 883 (1994)
  [gr-qc/9307009].

\bibitem{Camporesi:1995fb}
  R.~Camporesi and A.~Higuchi,
  J.\ Geom.\ Phys.\  {\bf 20}, 1 (1996)
  [arXiv:gr-qc/9505009].

\bibitem{Khanal:1981bb} 
  U.~Khanal and N.~Panchapakesan,
  Phys.\ Rev.\ D {\bf 24}, 829 (1981).
  
\bibitem{LopezOrtega:2012vi} 
  A.~Lopez-Ortega,
  Gen.\ Rel.\ Grav.\  {\bf 44}, 2387 (2012)
  [arXiv:1207.6791 [gr-qc]].

\bibitem{Chandrasekhar:1975zz} 
  S.~Chandrasekhar and S.~L.~Detweiler,
  Proc.\ Roy.\ Soc.\ Lond.\ A {\bf 345}, 145 (1975).

\bibitem{Crispino:2000jx} 
  L.~C.~B.~Crispino, A.~Higuchi and G.~E.~A.~Matsas,
  Phys.\ Rev.\ D {\bf 63}, 124008 (2001)
  [Erratum-ibid.\ D {\bf 80}, 029906 (2009)]
  [gr-qc/0011070].
  

\end{thebibliography}
\end{document}